\documentclass{ifacconf}

\usepackage{graphicx}      % include this line if your document contains figures
\usepackage{natbib}        % required for bibliography
\usepackage{amsmath,amssymb,xcolor}%,fullpage}
\usepackage{tikz}
\usetikzlibrary{decorations.pathreplacing}
\usepackage{algorithm}
\usepackage[noend]{algpseudocode}
\allowdisplaybreaks
\newcommand{\bi}{\begin{itemize}}
\newcommand{\ei}{\end{itemize}}
\newcommand{\vo}[1]{\textcolor[HTML]{000000}{\boldsymbol{#1}}}
\newcommand{\x}{\vo{x}}

\newcommand{\y}{\vo{y}}

\newcommand{\K}{\vo{K}}

\newcommand{\Q}{\vo{Q}}

\newcommand{\R}{\vo{R}}

\newcommand{\I}{\vo{I}}

\newcommand{\bS}{\vo{S}}

\newcommand{\C}{\vo{C}}
\newcommand{\A}{\vo{A}}
\newcommand{\B}{\vo{B}}
\newcommand{\n}{\vo{n}}
\newcommand{\sP}{\vo{P}}

\newcommand{\w}{\vo{w}}

\newcommand{\real}{\mathbb{R}}
\newcommand{\Exp}[1]{\mathbb{E}\left[#1\right]}

\newcommand{\mup}{\vo{\mu}^{-}}

\newcommand{\mupp}{\vo{\mu}^{+}}

\newtheorem{theorem}{Theorem}

\newcommand{\eqnlabel}[1]{\label{eqn:#1}}

\newcommand{\eqn}[1]{(\ref{eqn:#1})}

\newcommand{\Uarei}{\begin{align}
&\sP\A + \A^T\sP + \Delta\A^T\sP\A - (\Delta\A^T+\I)\sP\B\nonumber\\&\times({\I}+\Delta\B^T\sP\B)^{-1}\B^T\sP(\Delta\A+\I)+\Q=0,\eqnlabel{uarei}\end{align}}
\newcommand{\Uarer}{\begin{align}
&\sP\A + \A^T\sP + \Delta\A^T\sP\A - (\Delta\A^T+\I)\sP\B\nonumber\\&\times({\R}+\Delta\B^T\sP\B)^{-1}\B^T\sP(\Delta\A+\I)+\Q=0,\eqnlabel{uarer}\end{align}}

\begin{document}
\begin{frontmatter}

\title{Eigen Value Analysis in Lower Bounding Uncertainty of Kalman Filter Estimates\thanksref{footnoteinfo}}

\thanks[footnoteinfo]{This research was sponsored by Air Force Office of Scientific Research, Dynamic Data Driven Applications Systems grant FA9550-15-1-0071}

\author{Niladri Das \& Raktim Bhattacharya}

\address{Department of Aerospace Engineering,
        Texas A\&M University, College Station, Texas, USA. (e-mail: niladridas,raktim@tamu.edu).}

\begin{abstract}
In this paper we are concerned with the error-covariance lower-bounding problem in Kalman filtering: a sensor releases a set of measurements to the data fusion/estimation center, which has a perfect knowledge of the dynamic model, to allow it to estimate the states, while preventing it to estimate the states beyond a given accuracy. We propose a measurement noise manipulation scheme to ensure lower-bound on the estimation accuracy of states. Our proposed method ensures lower-bound on the steady state estimation error of Kalman filter, using mathematical tools from eigen value analysis.
\end{abstract}


\end{frontmatter}
%===============================================================================

\section{Introduction}
In various applications such as air traffic in \cite{1261143}, ground traffic  in \cite{work2008ensemble}, power grid in \cite{ghahremani2011dynamic}, and health monitoring in \cite{yi2013wavelet}, state estimation plays in important role. Kalman filtering covers a wide variety of such applications. A Kalman filter used in any of these scenarios provides information about the accuracy of the state estimates. For an air traffic system this might be the accuracy of the GPS location, whereas for the smart grid this might be the power consumption pattern for a house-hold. These estimates rely on the measurement data shared by the user itself and might be publicly available. Publicly available measurement data can be utilized with a knowledge about the dynamics to accurately estimate states of a particular individual entity such as a house-hold power consumption pattern or states of a covert satellite.

The primary concern in each of these cases is to ensure that the measurement data can not be used by someone with exact knowledge of the dynamics to estimate the states beyond certain accuracy. The optimal strategy to ensure this is, not sharing any measurement data at all. This strategy is impractical because a user who wants privacy might be a part of the bigger network whose operation relies on data sharing. For a smart grid reliable data sharing of the house-hold consumption ensure optimal power distributions, whereas in case of satellites this ensures reduced risk of collision between two active satellites which have undergone orbit changing maneuvers.

Ruling out the possibility of not sharing any data, the problem essentially convert to the following: \textit{how can we transform the measurement data that will ensure state estimation error, using a Kalman filter, of some or all of the states to be always above a certain accuracy level ?}

\noindent \textbf{Contributions of the paper:}
On a formulation level, if the system dynamics for states $\vo{x}$ is:
$\x_{k+1} = \A\x_{k} + \B\w_{k}, \ \forall k\in \mathbb{N},$ and the measurement equation for measurements $\vo{y}$ is:
$\y_k = \C\x_k + \n_k, \ \forall k\in \mathbb{N}, $ whose Kalman filtering based covariance update equation is:
\begin{align}
\vo{P}_{k|k} &= \vo{P}_{k|k-1}-\vo{P}_{k|k-1}\vo{C}^{T}(\C\vo{P}_{k|k-1}\vo{C}^{T}+\vo{R})^{-1}\nonumber\\&\times\vo{C}\vo{P}_{k|k-1}\nonumber
\end{align}
where $\vo{P}_{k|k-1}$ and $\vo{P}_{k|k}$ denotes the prior and posterior covariance matrix, the question that we are interested in answering is: manipulate $\vo{R}$ matrix such that a certain lower-bound is satisfied. Regulating $\vo{R}$ is akin to adding synthetic measurement noise, deciding noise intensity for active sensors such as lidar or laser, and scheduling or selecting sensing regimes. Sensor selection techniques are covered in \cite{singh2017supermodular},\cite{zhang2017sensor},and \cite{tzoumas2016near} among others.

Similar question is dealt most recently in \cite{Song_2018}. The authors investigate the existence of a linear transformation of the measurement space, compressing the measurement information, thus inflating the estimation error covariance. Apart from using a linear transform, we can also regulate the covariance of the steady state error estimates by adding noise to the measurements. This is the major difference between our work and the existing ones, is that we are interested in calculating the measurement noise covariance that satisfies the of lower-bound on the estimation error covariance.

%\begin{enumerate}
%\item Importance of manipulating sensor noise
%\item Addition of synthetic noise
%\item Importance of lower bounding the estimate covariance
%\item Give motivating examples
%\item \textcolor{red}{Talk about privacy angle of this problem}
%\item Talk about the SSA data sharing
%\item Talk about data sharing for internet of things
%\item Data sharing in advanced air traffic.
%\end{enumerate}

\noindent \textbf{Notations:}  Let $\mathbb{N}$ and $\real$ ($\real_{+}$) represent the sets of natural number and real (positive real) numbers respectively. The state space of system is a closed set in $\real^{n_x}$, where $n_x$ is the dimension of the states. Transpose of a square matrix $\mathbf{M} \in \real^{n\times n}$ is denoted as $\mathbf{M}^{T}$. A positive definite
(semi-definite) matrix $\mathbf{M}$ is denoted by $\mathbf{M}\succ 0$ ($\mathbf{M}\succeq 0$) and $\mathbf{M}\succ \boldsymbol{N}$ ($\mathbf{M}\succeq \boldsymbol{N}$) if
$\mathbf{M}-\mathbf{N}\succ 0 $ ($\mathbf{M}-\mathbf{N}\succeq 0 $), for some matrix $\mathbf{N}\in \real^{n\times n}$. The set of all positive definite (semi-definite) matrices of size $n\times n$ is denoted by $\bS^{++}_{n}$ ($\bS^{+}_{n}$). Let $\lambda_{i}(\mathbf{M})$ denotes $i^{\text{th}}$ eigen value of the matrix $\mathbf{M}$, when we arrange them as $\lambda_1(\mathbf{M})\geq \lambda_2(\mathbf{M})\geq ... \geq \lambda_n(\mathbf{M})$. Similarly, singular values $\sigma_{i}(\mathbf{M})$ of $\mathbf{M}$, are arranged in non-increasing order:
$\sigma_1(\mathbf{M})\geq \sigma_2(\mathbf{M})\geq ... \geq \sigma_n(\mathbf{M})$.
Let diag($\mathbf{a}$) denotes a diagonal matrix, with $\mathbf{a}$ as its diagonal elements. We assume that $\x\in\mathcal{X}$ is continuous and $\mu(\x)$ is a Lebesgue measure and p($\x$) is the probability density function (pdf). The expected value of the random variable $\x$ with respect to p($\x$) is represented as $\Exp{\x}$.

\noindent \textbf{Layout of the paper:}
The remainder of the paper is organized as follows. In Section 2, we present the system model along with its corresponding measurement model. In Section 3, we present Kalman filtering, leading to the problem statement in Section 4. In Section 5, we introduce preliminary results that lead us to the algorithm to solve the sensor covariances in section 6. In Section 7, the proposed framework is applied to a system. The paper finally concludes with Section 8.

\section{System dynamics and measurement model}
We focus on the class of \textit{discrete-time linear time invariant stochastic systems}. Let $\x_k$ represent the true states of a system at the $k^{\text{th}}$ time instant, where $\x_k \in \real^{n_{x}}$ for all $k\in\mathbb{N}$. The dynamics is modeled as:
\begin{equation}\x_{k+1} = \A\x_{k} + \B\w_{k}, \ \forall k\in \mathbb{N},   \eqnlabel{processDynamics}\end{equation}
where $\A\in \real^{n_{x}\times n_{x}}$ is the state transition matrix and matrix $\B \in \real^{n_{x}\times n_{w}}$. The process noise variable $\w_k \in\real^{n_w}$, is the  $n_w$ dimensional zero-mean Gaussian additive noise with $\mathbb{E}[\w_k\w_l^T] = \Q\delta_{kl}$.\\
The discrete dynamics in \eqn{processDynamics} is observed by a linear measurement model. Let $\y_k\in\real^{n_y}$ denote the measurement taken at the $k^{\text{th}}$ time instant as:
\begin{equation}\y_k = \C\x_k + \n_k, \ \forall k\in \mathbb{N}, \eqnlabel{sensing}\end{equation}
where $\y_k$ is corrupted by a $n_y$ dimensional additive observation noise  $\n_k \in\real^{n_n}$. The sensor noise at each time instant is a zero mean Gaussian random variable with $\mathbb{E}[\n_k\n_l^T] = \R\delta_{kl}$. The matrix $\C\in \real^{n_{y}\times n_{x}}$ is known as the observation or the measurement matrix.

The initial state of \eqn{processDynamics} is modeled by a Gaussian random variable $\x_0$ with mean $\boldsymbol{\mu}_{0}$ and  covariance $\mathbf{P}_0$. The random variable $\x_0$ denotes the system state at the $0^{\text{th}}$ time instant. The process noise $\w_{k}$, observation noise $\n_k$ and initial state variable $\x_0$ are all assumed to be independent, unless otherwise specified. These assumptions are strongly motivated by analytic tractability. The restriction to zero-mean noise sources is not a loss of
generality. When the noise sources are no longer zero mean, the $\A,\B$, and $\C$ matrices are modified and extra states are introduced as shown in \cite{anderson1979optimal}.

\section{Kalman filtering}
The discrete time system in \eqn{processDynamics} and \eqn{sensing} induces a Kalman filter, as the optimal state estimator, with dynamics:
\begin{align}
\K_k &= \mathbf{P}_k^-\C^T\Big[\C\mathbf{P}_k^-\C^T+\mathbf{R}\Big]^{-1} \tag*{(Kalman Gain)},\\
\mup_{k} & = \A\mupp_{k-1} \tag*{(Mean Propagation)},\\
\mathbf{P}_k^- &= \A\mathbf{P}^{+}_{k-1}\A^T+\B\Q\B^T \tag*{(Covariance Propagation)},\\
\mupp_{k}  &=\A\mupp_{k-1} + \K_k(\y_k-\C\mup_{k})\tag*{(Mean Update)},\\
\mathbf{P}^{+}_{k}&= (\mathbf{I}-\K_k\C)\mathbf{P}_k^-\tag*{(Covariance Update)},\\
\boldsymbol{\mu}^{+}_{0}&=\boldsymbol{\mu}_{0}\tag*{(Initial State Mean)},\\
\mathbf{P}^{+}_{0}&=\mathbf{P}_0\tag*{(Initial State Covariance)},
\end{align}
where the variables $\mup_{k},\mupp_{k}\in \real^{n_x}$, denotes the prior and posterior mean estimate of the random variable $\x_k$,  and $\boldsymbol{K}_k$ is the Kalman gain, at time $k$. The positive semi-definite matrices $\mathbf{P}_k^-,\mathbf{P}_k^+ \in \real^{n_x\times n_x}$ are the prior and posterior covariance matrix at time instant $k$ respectively.  Matrix inverse of the observation noise covariance $\mathbf{R}$ is defined as the \textit{precision matrix} $\bS$.

\section{Problem statement}
The problem that we address is as follows:
we assume that the system matrices ($\A,\B,\C$) and noise parameter $\Q$ of \eqn{processDynamics} and \eqn{sensing} are all known. The matrix $\R$ which is the sensor noise covariance, is the design variable. For a prescribed lower bound on the \textit{steady state prior covariance} of the state error estimate using Kalman filter, we need to design $\R$ or the precision matrix $\bS:=\R^{-1}$, that satisfies a prescribed \textit{lower-bound} on the steady state error covariance.

The final result is presented as a theorem. The proof of this theorem depends upon another theorems that we first prove in the succeeding section as preliminary results.
% \textcolor{red}{See if you can write an outline of the techniques you have used to arrive at the final theorem. This will give readers a motivation of why we need the succeeding two theorems}

\section{Preliminaries}
\subsection{Preliminary results:}

\cite{middleton1990digital} introduced the Unified Algebraic Riccati Equation :
\Uarei
where $\A\in\mathbb{R}^{n_{x}\times n_{x}}$ and $\B\in\mathbb{R}^{n_{x}\times n_{y}}$ represent constant matrices, $\Q \in\mathbb{R}^{n_{x}\times n_{x}}$ is in $\mathbb{S}^{+}_{n}$, the matrix $\sP \in\mathbb{R}^{n_{x}\times n_{x}}$ is the positive definite solution to \eqn{uarei}, and $\Delta$ represents sampling period.

\textbf{Note 1:} Matrix $\B$ in \eqn{uarei} is  not the same as in \eqn{processDynamics}.

We introduce an extra parameter $\R\in \mathbb{R}^{n_{y}\times n_{y}}$ in UARE and call it UARE-R. This UARE-R:
\Uarer is often encountered in Optimal Control and Estimation problems such as in \cite{bryson2018applied} and \cite{anderson1979optimal}.

\textbf{Remarks 1:} (a) Using $\Delta=0$, replacing $\A$ by $\A^{T}$, and $\B$ by $\C^T$, we recover the Continuous Time Algebraic Riccati Equation (CARE), solution to which gives us the steady state covariance for a  Kalman-Bucy filter. (b) Using $\Delta=1$, replacing $\A+\I$ by $\A^T$, and $\B$ by $\C^T$ we recover the Discrete Algebraic Riccati Equation (DARE) associated with steady state covariance of the Kalman Filter, where $\sP$ denotes the steady-state error covariance matrix.

Reiterating, our objective is to design $\R$ matrix that satisfies prescribed lower-bound on the steady-state estimated state error covariance matrix $\sP$, using \eqn{uarer}. We closely follow the calculations in \cite{lee2003matrix} to relate the bounds on $\R$ with that of $\sP$ in \eqn{uarer}.

\subsection{Eigen value based analysis}
In the following two theorems we examine the characterization of the bounds on the $\sP$ matrix of the UARE-R, as a function of $\R$.
As our final result we provide the theorem that connects the eigen values of $\R$ to lower bounds on $\sP$. This opens up a way to generate the feasible set for choosing $\R$ matrix. \\

\begin{theorem}\label{thm:neat5}
Let $\sP$ be the positive solution of the UARE-R in equation \eqn{uarer}, then
\begin{align}
\sP \succeq &(\Delta\A+\I)^T(\sP_{l0}^{-1}+\Delta\B\R^{-1}\B^T)^{-1} (\Delta\A+\I) \nonumber\\&+\Delta\Q\equiv \sP_{l1}
\end{align}
where the matrix $\sP_{l0}$ is defined as,
\begin{align}
\sP_{l0}\equiv&(\Delta\A+\I)^T(\varphi^{-1}\I+\Delta\B\R^{-1}\B^T)^{-1}(\Delta\A+\I)\nonumber\\&+\Delta\Q
\end{align}
and the positive constant $\varphi$ is defined as,
\begin{align}
\varphi\equiv &f(-[\lambda_{n_{x}}(\A+\A^T+\Delta\A^T\A)+\Delta\lambda_{n_{x}}(\Q)\lambda_1(\R^{-1})\nonumber\\&\times\sigma_1^2(\B)],2\lambda_1(\R^{-1})\sigma_1^2(\B),2\lambda_{n_{x}}(\Q)),
\end{align}
where $f(a,b,c)$ is defined as,
\begin{align}
f(a,b,c)\equiv \frac{-a+\sqrt{a^2+bc}}{b}.
\end{align}
\end{theorem}

\begin{pf}
We have,
\begin{align}\Delta(\sP\A + \A^T\sP + & \Delta\A^T\sP\A)\nonumber\\= &(\Delta\A+\I)^T\sP(\Delta\A+\I)-\sP, \eqnlabel{eq1}\end{align}
Re-writing UARE-R as:
\begin{align}&\Delta(\sP\A + \A^T\sP + \Delta\A^T\sP\A) - \Delta(\Delta\A^T+\I)\nonumber\\&\times\sP\B({\R}+\Delta\B^T\sP\B)^{-1}\B^T\sP(\Delta\A+\I)+\Q=0,
\end{align}
and using \eqn{eq1},
\begin{align}
&(\Delta\A+\I)^T[\sP-\Delta\sP\B({\R}+\Delta\B^T\sP\B)^{-1}\B^T\sP]\nonumber\\&\times(\Delta\A+\I)+\Delta\Q=\sP.  \eqnlabel{eq2}
\end{align}
Using Matrix Inversion lemma we get,
\begin{align}
&(\Delta\A+\I)^T(\sP^{-1}+\Delta\B\R^{-1}\B^T)^{-1} (\Delta\A+\I)\nonumber\\&+\Delta\Q=\sP.\eqnlabel{uaremod}
\end{align}
Following \cite{lee2003matrix} we have,
\begin{align}
\sP^{-1} &\preceq \lambda_{n_{x}}^{-1}(\sP)\I.\eqnlabel{eq3}\\
\B\R^{-1}\B^T &\preceq \lambda_1(\R^{-1})\sigma_1^2(\B)\I.\eqnlabel{eq4}
\end{align}
Using \eqn{eq3} and \eqn{eq4} in \eqn{uaremod}, we have:
\begin{align}
\sP&\succeq\frac{\lambda_{n_{x}}(\sP)}{1+\lambda_{n_{x}}(\sP)\Delta\lambda_1(\R^{-1})\sigma_1^2(\B)}(\Delta\A+\I)^T\nonumber\\&\times(\Delta\A+\I)+\Delta\Q\eqnlabel{eq6}
\end{align}
Lemma 1 in \cite{lee2003matrix} (Amir-Moez 1956) states,
\begin{align}
\lambda_{i+j-n}(\A_{1}+\B_{1}) \geq \lambda_j(\A_{1})+\lambda_i(\B_{1}), \ i+j \geq n+1,\nonumber
\end{align}
for any symmetric matrices, $\A_{1},\B_{1}\in\mathbb{R}^{n\times n}$ and $1\leq i,j \leq n$.\\
Using $i=n$ and $j=n$ we have,
\begin{align}
\lambda_{n}(\A+\B) \geq \lambda_n(\A_{1})+\lambda_n(\B_{1}).\eqnlabel{eq7}
\end{align}
Using equation \eqn{eq7} in equation \eqn{eq6} after applying eigen value operator on equation \eqn{eq6},  we get,
\begin{align}
\lambda_{n_{x}}(\sP)&\geq\frac{\lambda_{n_{x}}(\sP)}{1+\lambda_{n_{x}}(\sP)\Delta\lambda_1(\R^{-1})\sigma_1^2(\B)}\lambda_{n_{x}}[(\Delta\A+\I)^T\nonumber\\&\times(\Delta\A+\I)]+\Delta\lambda_{n_{x}}(\Q).\nonumber
\end{align}
Using $\lambda_{n_{x}}[(\Delta\A+\I)^T(\Delta\A+\I)]=\Delta\lambda_{n_{x}}(\A+\A^T+\Delta\A^T\A)+1$ and then rearranging we get,
\begin{align}
&\Delta\lambda_1(\R^{-1})\sigma_1^2(\B)\lambda_{n_{x}}^2(\sP)-[\Delta\lambda_{n_{x}}(\A+\A^T+\Delta\A^T\A)\nonumber\\&+\Delta\lambda_{n_{x}}(\Q)\Delta\lambda_1(\R^{-1})\sigma_1^2(\B)]\lambda_{n_{x}}(\sP)-\Delta\lambda_{n_{x}}(\Q) \geq 0,\eqnlabel{eq8}
\end{align}
which is of quadratic form.
Hence finally,
\begin{align}
&\lambda_{n_{x}}(\sP)\geq f(-[\lambda_{n_{x}}(\A+\A^T+\Delta\A^T\A)+\Delta\lambda_{n_{x}}(\Q)\nonumber\\&\times\lambda_1(\R^{-1})\sigma_1^2(\B)],2\lambda_1(\R^{-1})\sigma_1^2(\B),2\lambda_{n_{x}}(\Q))\equiv \varphi,\nonumber
\end{align}
where,
\begin{align}
f(a,b,c)\equiv \frac{-a+\sqrt{a^2+bc}}{b}.
\end{align}
We have,
\begin{align}
\lambda_n^{-1}(\sP) \leq & \ \varphi^{-1}\eqnlabel{eq10}
\end{align}
Using equation \eqn{eq10} in equation \eqn{eq6} we get,
\begin{align}
\sP&\succeq (\Delta\A+\I)^T(\varphi^{-1}\I+\Delta\B\R^{-1}\B^T)^{-1}\nonumber\\&\times(\Delta\A+\I)+\Delta\Q \equiv \sP_{l0}
\end{align}
Using the lower bound $\sP_{l0}$ in \eqn{uaremod} we get,
\begin{align}
\sP &\succeq (\Delta\A+\I)^T(\sP_{l0}^{-1}+\Delta\B\R^{-1}\B^T)^{-1} \nonumber\\&\times(\Delta\A+\I)+\Delta\Q
\end{align}

\end{pf}

Theorem \ref{thm:neat5} is similar to that of Theorem 1 of \cite{lee2003matrix} but derived for UARE-R, rather than for UARE.

\section{Choosing $\vo{R}$ to Lower-bound $\sP$}
We discussed how we retrieve DARE:
\begin{align}
&\A\sP\A^T-\sP-\A\sP\C^T(\R+\C\sP\C^T)^{-1}\nonumber\\&\times\C\sP\A^T+\Q=0,
\nonumber
\end{align}
 for solving the steady-state covariance matrix $\sP$ for Kalman filter, applying suitable substitution to the UARE-R. Conventionally, designing $\R$ is related to upper bounding the performance of a filter with some additional constraints. Topics like differential privacy as in \cite{dwork2014algorithmic} and bounded information exchange such as in robotics as in \cite{butler2015privacy} has lead to the requirement of switching between different  $\R$ matrices to keep the performance within bounds (upper or lower), rather than just upper bounding it. In this work, we utilize Theorem \ref{thm:neat5}  to propose a technique to design the measurement noise covariance matrix $\R$ or the precision matrix  $\bS$ such that the $\sP$ is lower bounded. We will see in the succeeding sections that the feasible set of $\R$ is represented as a set of linear matrix inequality (LMI).

 % In the following results the $\R$ matrix or the $\bS:=\R^{-1}$ is assumed to be a diagonal matrix. Hence the spectrum of $\bS$ consists of its diagonal elements. We use $\vo{\lambda}$ to represent the diagonal of $\bS$.

In deriving the following result, we first construct the feasible set of $\R$ that satisfies prescribed lower bound on the matrix $\sP$. A particular choice of $\R$ matrix results from an optimization problem over the set of feasible $\R$ for a given cost function. We use $c(\R)$ to represent a generic cost function.

\textbf{Remarks 2:} If $\R$ is a diagonal matrix, the cost function is essentially over the space of vector $\vo{\lambda}$, that constitutes the diagonal elements of $\R$.

\subsection{Calculate $\vo{R}$: Lower bound for steady state Kalman filtering}
Using $\Delta=1$, replacing $\A+\I$ by $\A^T$ and $\B$ by $\C^T$ in UARE-R, we recover the Discrete Algebraic Riccati Equation (DARE) associated with steady state covariance update equation of a linear system using Kalman Filter.
%
%We use the same transformations on Theorem \ref{thm:neat7} and \ref{thm:neat8} to derive results for Kalman Filter.

We assume complete detectability of [$\A,\C$] and stabilizability of [$\A,\B\Q^{1/2}$] (\cite{anderson1979optimal}, pg.82) for \eqn{processDynamics} and \eqn{sensing}. This ensure that the steady state prior covariance matrix $\sP$ exists and is unique (for a fixed $\R$) for the corresponding DARE.

\begin{theorem}\label{thm:neat7}
For a given scalar cost function c($\boldsymbol{R}$) and an lower bound $(1/\lambda^f_u)$ on the spectrum of $\R$, the solution $\R^*$, whose spectrum is $$\lambda(\R^*):=\{\lambda_{1}\geq\lambda_2\geq\dots\geq\lambda_{ny-1}\geq\lambda_{ny}\},$$ where $\lambda_{ny}\geq (1/\lambda^f_{u})$, that satisfies a given lower bound $\sP_l^f$ on the steady state prior covariance matrix $\sP$ of Kalman filter, is given by the following optimization problem.
\begin{align}
\R^*&:=\underset{\R}{\mathrm{argmin}} \  c(\R)\nonumber
\end{align}
Such that,
\begin{align}
\R &\succeq (1/\lambda^f_{u})\I,\nonumber\\
\begin{bmatrix}\vo{T}_{1} & \C^{T} \\ \C & \R\end{bmatrix}&\succeq 0,\nonumber
\end{align}
where,
\begin{align}
\vo{T}_{1}\equiv&\A^{T}(\sP_l^f -\Q)^{-1}\A -{\sP^{'}_{l0}}^{-1}\nonumber\\
\sP^{'}_{l0} \equiv & \A(\varphi'^{-1}\I+\lambda^f_{u}\C^{T}\C)^{-1}\A^{T}+\Q.\nonumber\\
\varphi'\equiv &f(-[\lambda_{n_{x}}(\A\A^{T}-\I)+\lambda_{n_{x}}(\Q)\lambda^f_{u}\sigma_1^2(\C^T)]\nonumber\\&,2\lambda^f_{u}\sigma_1^2(\C^T),2\lambda_{n_{x}}(\Q)),\nonumber
\end{align}
\end{theorem}
\begin{pf}
We first take a look at Theorem \ref{thm:neat5}, the lower bound theorem. The variable $\varphi$ is defined as,
\begin{align}
\varphi&\equiv f(-[\lambda_{n_{x}}(\A+\A^T+\Delta\A^T\A)+\Delta\lambda_{n_{x}}(\Q)\lambda_1(\R^{-1})\nonumber\\&\times\sigma_1^2(\B)],2\lambda_1(\R^{-1})\sigma_1^2(\B),2\lambda_{n_{x}}(\Q))\nonumber
\end{align}
We notice that $\varphi$ is a function of $\lambda_1(\R^{-1})$. We assume that $\lambda_1(\R^{-1})\leq \lambda^f_{u}$, i.e. upper bounded. We define:
\begin{align}
\varphi'&\equiv f(-[\lambda_{n_{x}}(\A+\A^T+\Delta\A^T\A)+\Delta\lambda_{n_{x}}(\Q)\lambda^f_{u}\nonumber\\&\times\sigma_1^2(\B)],2\lambda^f_{u}\sigma_1^2(\B),2\lambda_{n_{x}}(\Q))\nonumber
\end{align}
It can be proved that $\varphi\geq \varphi'$.
Hence we have,
\begin{align}
\sP_{l0}\succeq&(\Delta\A+\I)^T(\varphi^{-1}\I+\Delta\lambda_1(\R^{-1})\B\B^T)^{-1}\nonumber\\&\times(\Delta\A+\I)+\Delta\Q,\nonumber\\
\succeq&(\Delta\A+\I)^T(\varphi'^{-1}\I+\Delta\lambda^{f}_{u}\B\B^T)^{-1}\nonumber\\&\times(\Delta\A+\I)+\Delta\Q\equiv \sP^{'}_{l0},
\end{align}
Using Theorem \ref{thm:neat5} we have,
\begin{align}
\sP &\succeq (\Delta\A+\I)^T(\sP_{l0}^{-1}+\Delta\B\R^{-1}\B^T)^{-1}\nonumber\\&\times (\Delta\A+\I)+\Delta\Q\nonumber\\
&\succeq (\Delta\A+\I)^T({\sP'_{l0}}^{-1}+\Delta\B\R^{-1}\B^T)^{-1} \nonumber\\&\times(\Delta\A+\I)+\Delta\Q.\nonumber
\end{align}
Now suppose we want to lower bound $\sP$ by $\sP_l^f$. That is ensured if we have,
\begin{align}
\sP \succeq &(\Delta\A+\I)^T({\sP'_{l0}}^{-1}+\Delta\B\R^{-1}\B^T)^{-1} \nonumber\\&\times(\Delta\A+\I)+\Delta\Q \succeq \sP_l^f.\nonumber
\end{align}
If we assume that the $(\Delta \A+\I)$ matrix is invertible we have,
\begin{align}
(\Delta\A+\I)^{-T}&(\sP_l^f-\Delta\Q)(\Delta\A+\I)^{-1} \nonumber\\&\preceq ({\sP'_{l0}}^{-1}+\Delta\B\R^{-1}\B^T)^{-1} \nonumber
\end{align}
Since ${\sP'_{l0}}^{-1}\succeq 0$, $({\sP'_{l0}}^{-1}+\Delta\B\R^{-1}\B^T)$ is invertible, we have,
\begin{align}
(\Delta\A+\I)&(\sP_l^f -\Delta\Q)^{-1}(\Delta\A+\I)^{T} \nonumber\\&\succeq ({\sP'_{l0}}^{-1}+\Delta\B\R^{-1}\B^T) \nonumber\\
(\Delta\A+\I)&(\sP_l^f -\Delta\Q)^{-1}(\Delta\A+\I)^{T} -{\sP'_{l0}}^{-1} \nonumber\\&\succeq \Delta\B\R^{-1}\B^T=\Delta\B\boldsymbol{S}\B^T\label{leq1}.
\end{align}
For $0\leq\lambda_1(\boldsymbol{S})\leq \lambda^{f}_{u}$ and given lower bound on $\sP$, which is $\sP_l^f$, we can calculate the feasible solutions to the diagonal $\boldsymbol{S}$ matrix. Using $\Delta=1$, replacing $\A+\I$ by $\A^T$ and $\B$ by $\C^T$ we get:
\begin{align}
\C^{T}\R^{-1}\C &\preceq \A^{T}(\sP_l^f -\Q)^{-1}\A -{\sP^{'}_{l0}}^{-1},\nonumber\\
\C^{T}\R^{-1}\C &\preceq \vo{T}_{1}\nonumber\\
\begin{bmatrix}\vo{T}_{1} & \C^{T} \\ \C & \R\end{bmatrix}&\succeq 0, \nonumber
\end{align}
Where $\vo{T}_{1}\equiv \A^{T}(\sP_l^f -\Q)^{-1}\A -{\sP^{'}_{l0}}^{-1}$
\end{pf}

\subsection{Choosing feasible lower bound of $\sP$ for Kalman filter}
The desired covariance bounds on $\sP$ should be chosen carefully.
When system matrices $\A,\B,\C$ and noise parameter $\Q$ is already chosen or are known, there exists an upper bound and lower bound on the $\sP$ for any choice of the matrix $\R$ under certain conditions. Choosing any positive definite matrices, as the desired $\mathbf{P}^f_{l}$, outside this bounds, will result in an infeasible solution for the $\R$ matrix. Hence it is important to choose the desired performance bound accordingly.
The prescribed $\mathbf{P}^f_{l}$ should lie between $\mathbf{P}^{lb}$ and $\mathbf{P}^{ub}$ satisfying the following:
\begin{align}
\mathbf{P}^{lb}&:=\A(\mathbf{P}^{lb}-\mathbf{P}^{lb}\C^T\Big[\C\mathbf{P}^{lb}\C^T\Big]^{-1}\C\mathbf{P}^{lb})\A^T+\B\Q\B^T\nonumber
\end{align}
The matrices $\mathbf{P}^{lb}$ is calculated using $\R=\vo{0}$ in the DARE. When $\R=\vo{0}$, the DARE is solved using generalized Shur method  as in \cite{articlesima} on an extended matrix pencil. The covariance $\mathbf{P}^{ub}$ satisfies the following:
\begin{align}
\mathbf{P}^{ub}&:= (\A\mathbf{P}^{ub}\A^T+\B\Q\B^T)\nonumber
\end{align}
The matrix $\mathbf{P}^{ub}$ is calculated by using $\R=\vo{\infty}$ in the DARE. An unique $\mathbf{P}^{ub}$ exists if $\A$ is stable.

In the succeeding section we apply our sensor $\R$ selection algorithm for a prescribed lower-bound on the steady state error covariance matrix.

\section{Numerical Expriment}
The system considered here is a $n_x$ dimensional discrete time linear Gaussian system. The $\B$ matrices are chosen to be identity.
The $\Q$ matrix is $\vo{I}$. The $\A$ and $\C$ matrices are chosen such that
[$\A,\C$] pair is detectable and [$\A,\B\Q^{1/2}$] pair is stabilizable. We choose $\R$ to be a diagonal matrix. Hence, the spectrum of $\R$, i.e. $\{\lambda_{i}\}$ are its diagonal elements.
We choose Theorem \ref{thm:neat7} and show results for \textit{minimizing} $l_1$ norm on $\vo{\lambda}$ ($\R:=\text{diag}(\vo{\lambda})$), for a prescribed lower bound on $\sP$, where $n_x=10$ and $n_y=10$. The matrix $\C$ in this example is chosen to be $2\I$. The matrices $\mathbf{P}^{lb}$ and $\mathbf{P}^{ub}$ are first calculated. The eigen values of
\begin{align}\text{eig}(\mathbf{P}^{ub})=[&
                    1.000 \
          1.001 \
           1.012 \
          1.123 \
          1.186 \
          2.139 \
          3.172 \nonumber\\
          &4.705 \
          9.096 \
          279.143],\nonumber
                  \end{align}
while the eigenvalues of $\mathbf{P}^{lb}$ all are equal to 1.

We then select the prescribed lower bound $\sP^{f}_{l}$ to be $(1/16)\times(\mathbf{P}^{ub}+15\mathbf{P}^{lb})$. This convex combination ensures a smooth transition from $\mathbf{P}^{lb}$ to $\mathbf{P}^{ub}$ when $\R$ goes from $\vo{0}$ to $\infty$.
We calculate $\varphi'=1.0000193$ and $\sP'_{u0}$.
We select the upper bound $\lambda_u^f$ to be 0.03.

The eigen values of $\sP'_{l0}$ :
\begin{align}\text{eig}(\sP'_{l0})=[&
                    28.689 \
          2.601 \
          2.028 \
          1.599 \
          1.480\nonumber\\
          &1.103 \
          1.078 \
          1.006 \
          1.000 \
          1.000].\nonumber
                  \end{align}

We solve the optimization problems using CVX in Matlab. The minimum $l_1$ norm cost is 18336.433 . On a 2GHz Intel Core i5 machine, the $l_1$ problem takes 1.20 seconds. 
%Since $\vo{\lambda}\in \real^{n_y}_{+}$, the $l_1$ norm minimization problem of $\vo{\lambda}$ is cast as minimization of trace of $\R=\text{diag}(\vo{\lambda})$.
\begin{figure}[!h]
\centering
\includegraphics[width=0.47\textwidth]{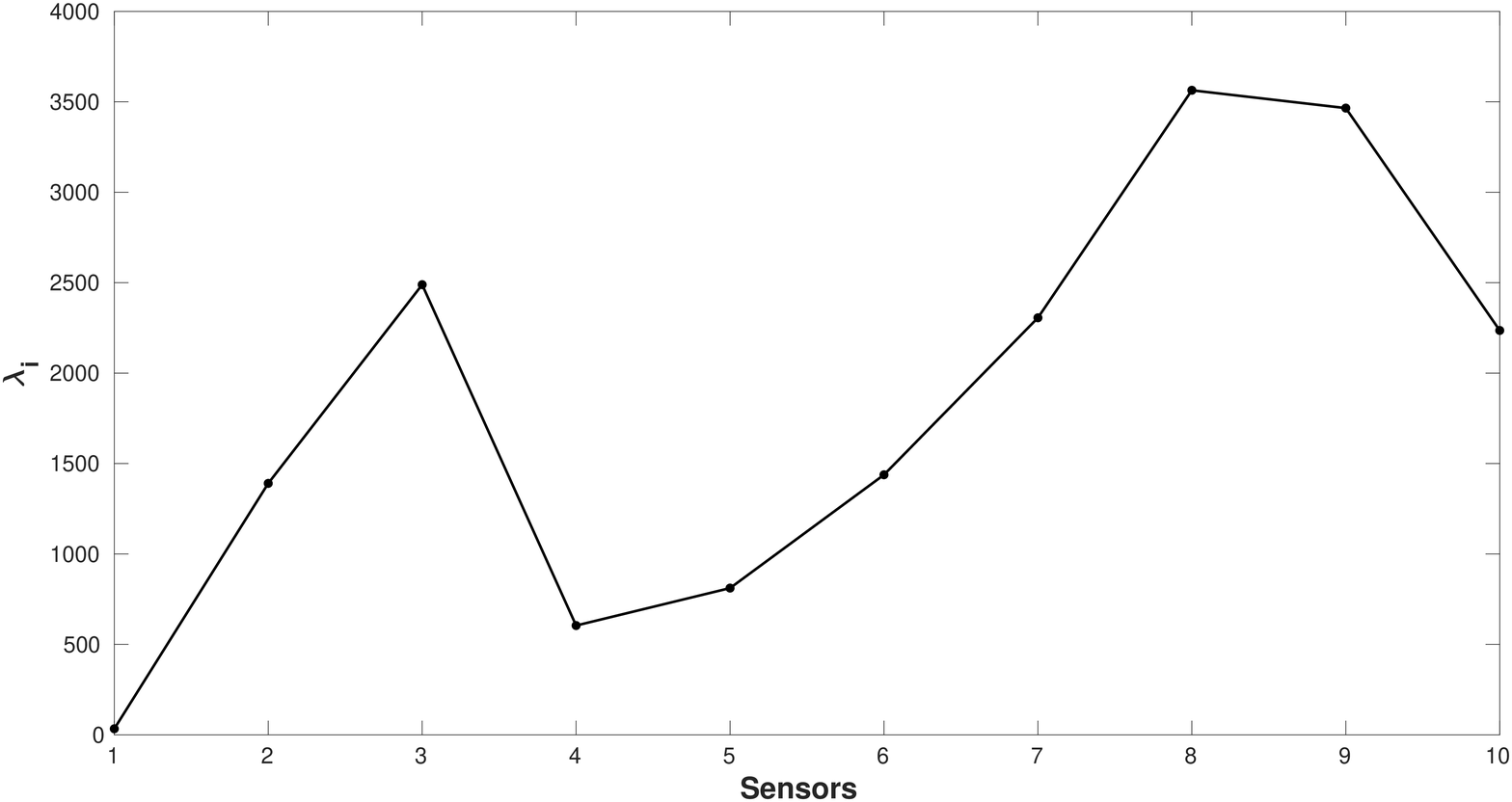}
\caption{Plot of sensor covariance values for 10 sensors for prescribed lower bound on $\sP$. Circle denotes covariance values calculated from minimization of $l_1$ norm of the vector $\vo{\lambda}$}
\label{unknown2}
\end{figure}

The $\vo{\lambda}$ vector of sensors for $l_1$ norm shown in fig(\ref{unknown2}) are:
\begin{align}    \vo{\lambda} =&          [
                    33.333 \
          1390.292 \
          2488.968 \
          604.108 \
          811.228 \nonumber\\
          &1437.797 \
          2305.879 \
          3563.793 \
          3465.244 \
          2235.787]\nonumber
        \end{align}

The $\R:=\text{diag}(\vo{\lambda})$ noise covariance that we calculated is the minimum noise that needs to be in the measurements to ensure that the steady state error covariance matrix is greater than the prescribed lower bound $\sP^{f}_{l}$. This is verified by calculating the eigen values of the $\sP-\sP^{f}_{l}$ matrix, which turns out to be all positive. Matrix $\sP$ is the DARE solution for the optimal $\R$. We notice that there is a large gap between the lower bound and the final steady state value of $\sP$. This is due to the fact that we used eigen value approximations in deriving the result. An ad-hoc method to reduce this gap is to iteratively reduce the magnitude of the $\vo{\lambda}$ till the eigenvalues of $\sP-\sP^{f}_{l}$ remain all positive. We found out that we can reduce the $\vo{\lambda}$ by a factor of $0.08$ and still ensure $\sP\succeq\sP^{f}_{l}$.

The calculated $\R$ can be assumed to be comprised of actual measurement noise due to the system $\R_a$ and synthetic noise $\R_s$. Since in most practical cases $\R_a$ is known, our algorithm effectively calculates the minimum synthetic noise $\R_s$ that needs to be added to the actual measurement to ensure privacy with respect to state estimation. 
%It is worth noting that the precision requirements of sensor 4, sensor 7, sensor 9, and sensor 10 are negligible.
%
%\section{Connections to Sensor Selection Problem}
%When $\sP$ is lower bounded, the noise covariances required by the sensors are shown in  \eqn{l2answer}. Four of the sensors (4,7,9,10) which have almost zero precision requirement can be safely removed. If we remove these sensors the resulting $\C$ matrix still preserves completely detectability of pair $(\A,\C)$. In fig.(\ref{unknown2}), we apply the algorithm to a larger set of 30 sensors, and we see that out of 30 sensors only five sensors have considerable sensor precision requirements. If we remove all the sensors except these 5, the modified $\C$ matrix still provide complete detectability. Hence we can safely remove the 25 sensors from the measurement model, and still satisfy the lower bounded on $\sP$.
%
%However, this pruning is ad hoc and it is possible that removal of some sensor with low precision may lead to loss of detectability. However, an iterative algorithm can be devised which eliminates the least precise sensor until there is loss of detectability. We can also apply the various algorithms to eliminate sensors, once their precisions are known.

\section{Conclusion}
In this paper we formulate an algorithm to calculate the measurement noise covariance which ensures that the steady state error covariance of the state estimates are lower-bounded by a prescribed bound. We introduce a modified  Unified Algebraic Riccati Equation (R-UARE) and exploit eigen value analysis to construct a feasible set of measurement noise covariance. This feasible set is convex and is represented as an LMI. We choose a convex cost function
over this convex feasible set of measurement noise covariance, and calculate an optimal noise covariance for system design, which is studied numerically in an example.
\bibliography{IEEEexample}

\end{document}